G. V. Ionov

# THE DETERMINATION OF THE TRAJECTORY OF CHELYABINSK BOLIDE ACCORDING TO THE RECORDS OF THE DRIVE CAMS AND THE SIMULATION OF THE FRAGMENTS MOTION IN THE ATMOSPHERE

*The determination of the trajectory of Chelyabinsk bolide according to the video records is performed and the results of the simulation of the fragment motion in the atmosphere are showed including its state at that moment. The methods of distortion compensation and adjusting of the video images with the calibration images and the iterative method of the trajectory improvement by azimuths and altitudes are developed. These methods allow improving the precision of the trajectory tracing to the hundreds of meters in space and to the tens of arc minutes in the angular measure.*

**Key words:** meteoroid**,** bolide, meteorite, fragment, video record, trajectory, ablation, head drag coefficient, azimuths, altitude, speed.

The characteristic of Chelyabinsk bolide of February 15, 2013 is its detection from the different locations of the ground surface that are located at the distances of several hundred kilometers from the place of event. It became possible due to the fair weather that was over the territory on February 15. The bolide was observed at the moment of sun rising and there was slight dusk. There were a lot of people in their cars with the operating video recorders this working morning. The background of the morning sky was sufficiently dark so the bolide trail that was lightened by the sunrays looked rich in contrast and view. The rising sun was caught by some video records and served as a guidemark in the angle measurement.

Due to the abundance of data on Chelyabinsk bolide the accurate model of the accompaniments can be built. These basic data are primarily the video records from the drive recorders. The video record from the video camera of Melnikov about the fall of the meteorite in the Chebarkul Lake supports the information about the final segment of moving of Chebarkul fragment. The characteristics of the formed ice crater, the places of the meteorite found and its characteristics can also serve for the calibration of the wind deflection model. There are a lot of photographs of the aerosol trace that was left by the bolide. These photographs show the time evolution of the trace and give the possibility to study the convection phenomena and transferring in the upper atmosphere that was done in the work of Nick Gorkavyi [1].

This work is dedicated primarily to the determination of the precise trajectory in all its segments. The trajectory undoubtedly determines the characteristic of the disintegration history of the original meteoroid, ablation processes, slowdown processes that lead to the transformation of kinetic energy of the meteoroid into the heat and optical radiation. The distribution of power along the trajectory that can be defined due to the profiles of illumination of the landscape details and the atmosphere flash is the basis for the calculation of the shock wave travel that caused the damage to Chelyabinsk and other nearest inhabited localities.

The base of the process model that went with the bolide is the trajectory, i.e., dependence on the time of location of the main body in space. It shall be named as the base fragment. At the moment of atmosphere entry it was like an original meteoroid and at the moment of falling it was like the Chebarkul meteorite. The following method was developed for the trajectory calculation. Primarily, the position in space of any one detail of trace is determined with the maximum precision. This pivot detail shall be selected as easy identifiable on many video records. The wind that is blowing in the upper atmosphere can reach the speed of tens of meters per second so the drift of the trace in a few seconds can be fairly noticed on the high resolution video records. Therefore, the identification of the pivot detail shall be performed in the first few seconds after the falling.

As for the further simulation it shall be considered that the trajectory goes through the position of the pivot detail in space. It remains to be found the direction along which the base fragment was moving. It shall be done in iterated way by the rotation of the trajectory in the space that leads it to the straight sight lines along which the luminous part of the bolide was observed from the different points at the ground surface. Whereas, the trajectory is set not in the form of the straight line but every time in case of direction changing, the full recalculating of the base fragment motion in the atmosphere shall be done including all physical processes.

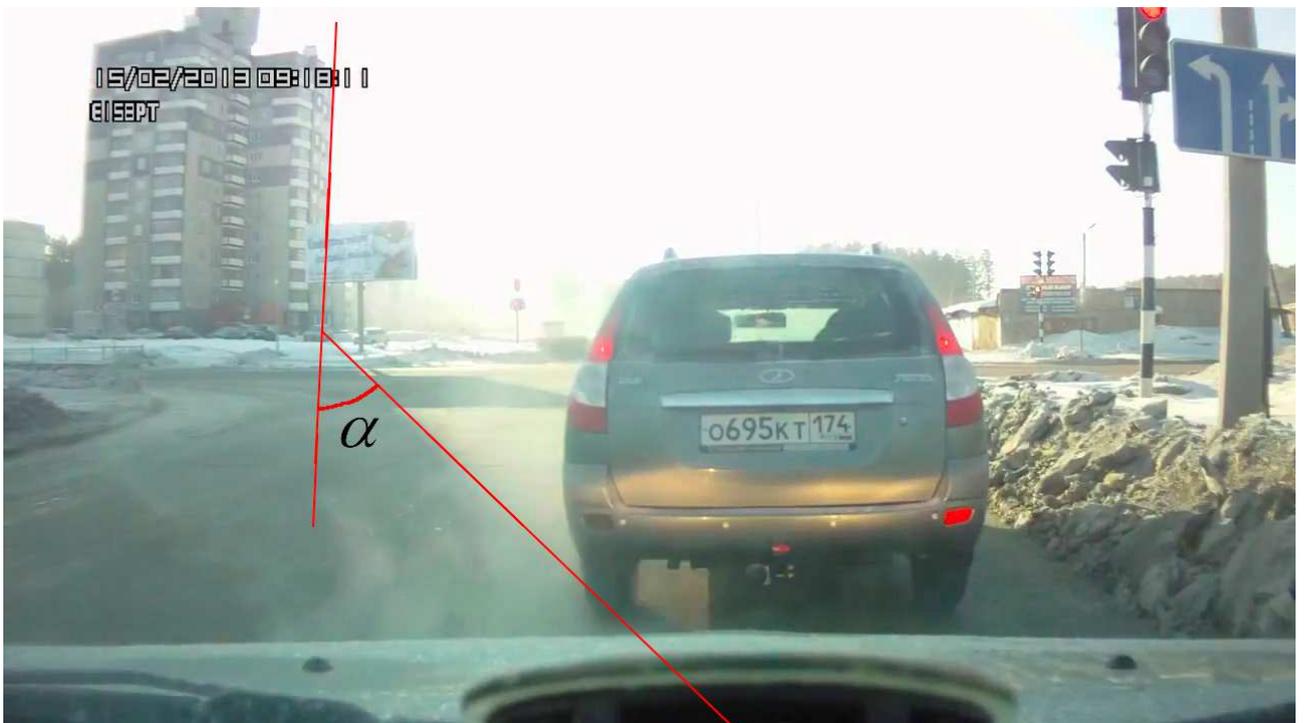

Figure 1. The shadow from the arris is on the horizontal road surface. The angle α is sensitive to the azimuth of the bolide and takes on a zero value at the moment of conjunction of azimuths of the bolide and the arris.

In order to determine the velocity of travel at the initial segment of the entry in the atmosphere, the measurements are used at the junction of streets of Lominskiy and Zababahin in the city of Snezhinsk. The video record from this junction [2] contains the moving of the shadows (the example of shadows on one of the images is shown in the figure 1) from the vertical arris of the tower building at the moments shadowing of the bolide and its moving out of the building that allows to find two time points for the known azimuths with the high precision (see figure 2, 3). If the trajectory position is known this will give us an opportunity to determine the velocity with the insignificant error that is specified by the indetermination of the trajectory position in the space but not by the measurement error on the video record from the city of Snezhinsk.

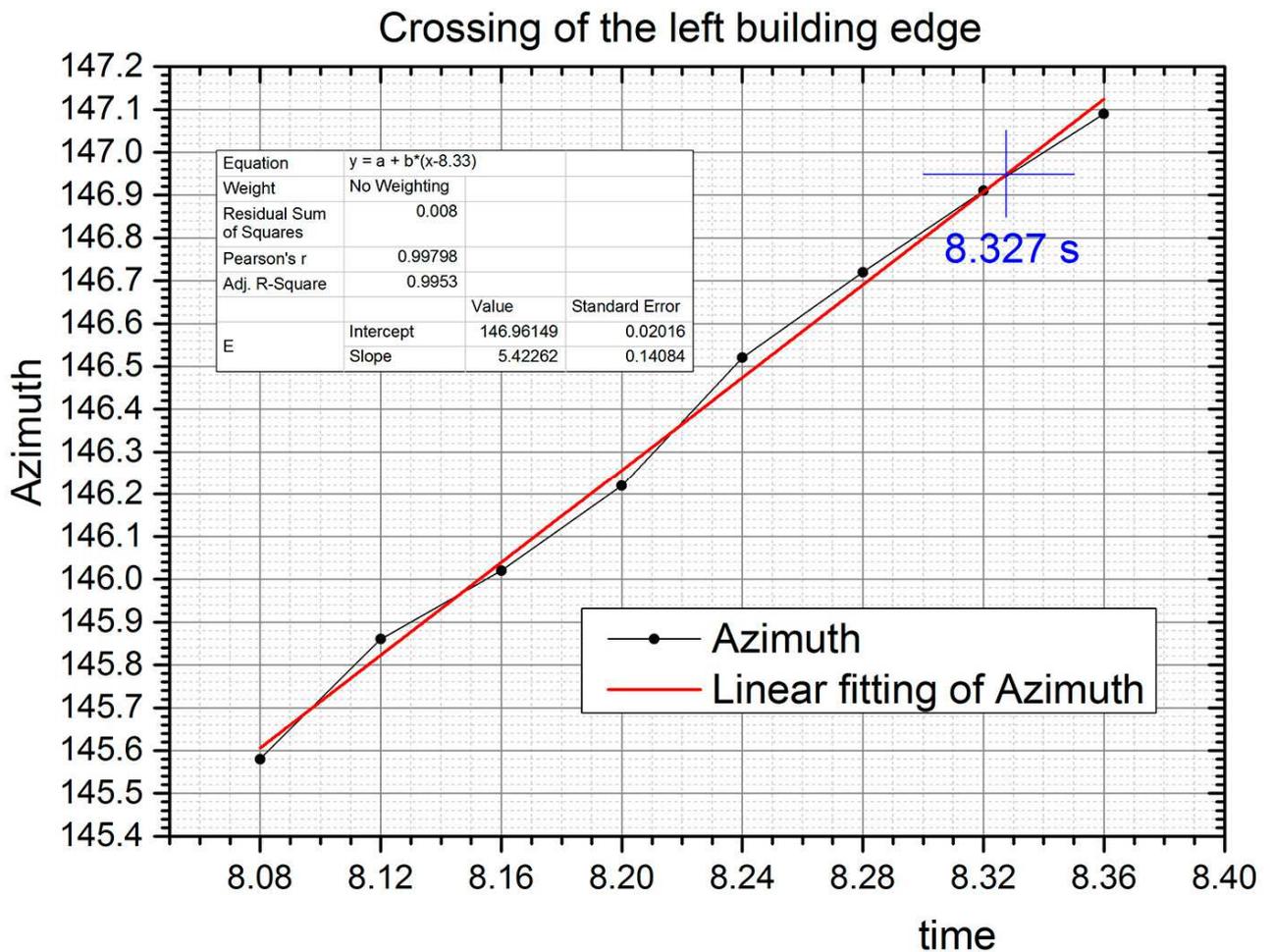

Figure 2. The dependence of azimuth of the base fragment on the time on video record from Snezhinsk [1], that is specified by α angle between the shadow and the arris of the building. The video record has the sequence of 25 frames per second, i.e. between the frames is 0.04 s. The sensitivity of α angle to the azimuth allows to increase the precision of the determination of the time moment when the bolide crosses the left arris of the building up to 0.007 s. The crossing was at 09:18:08.327 according to the timer on video record.

Due to the interference effect between the fragments through the shock waves, the direction of movement of the flown out fragments could slightly change during the main flash. Furthermore, due to the asymmetry of the forms and rotation, the lateral forces could affect the fragments (particularly, due to the Magnus effect). So it is seen in the video records that the fragments flow apart slightly (figure 4a).

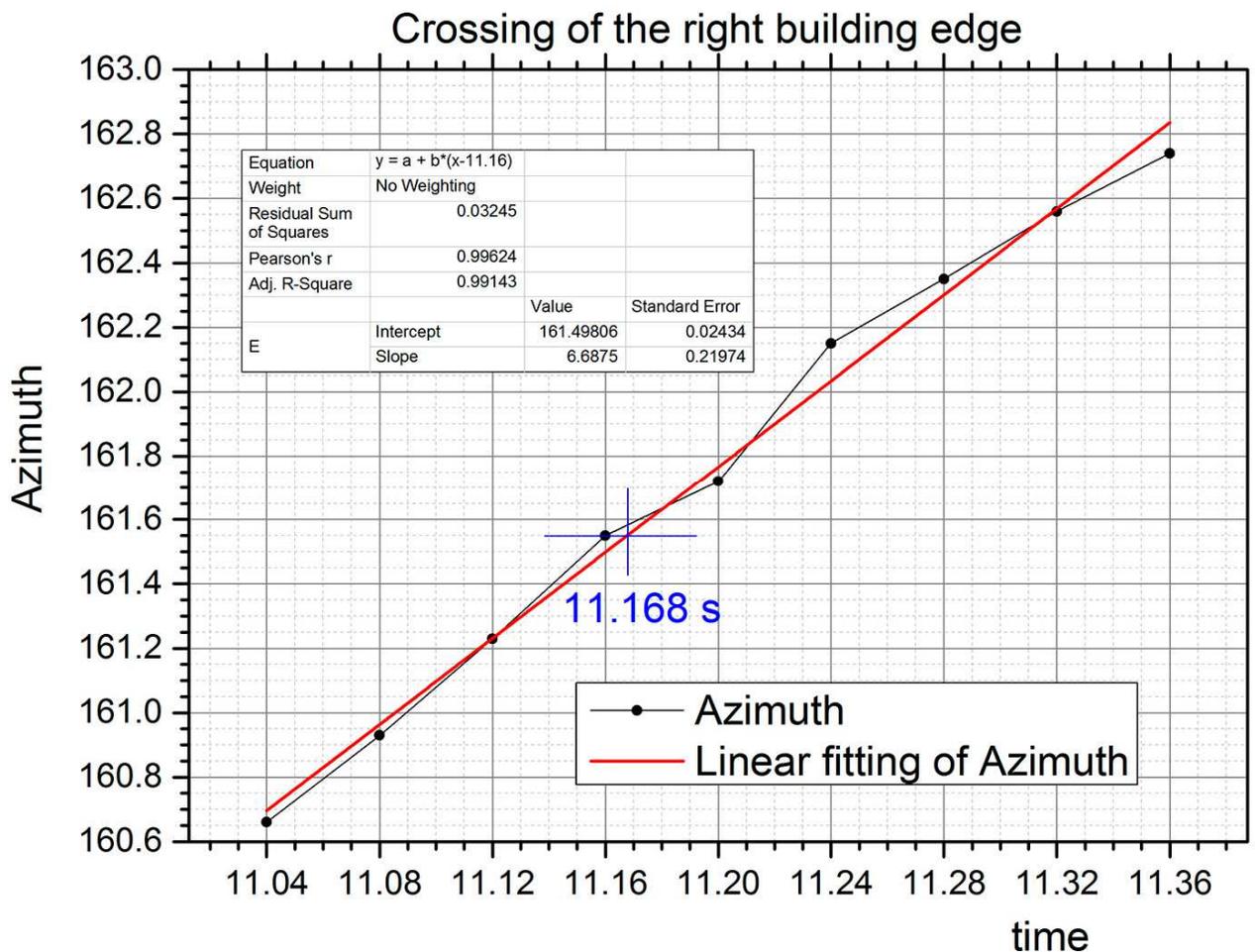

Figure 3. The dependence of azimuth of the base fragment on the time on video record from Snezhinsk [1], that is specified by α angle between the shadow and the arris of the building. According to the timer on video record the crossing of the right arris of the building occurred at 09:18:11.168. The error of the time determination is about 0.007 s.

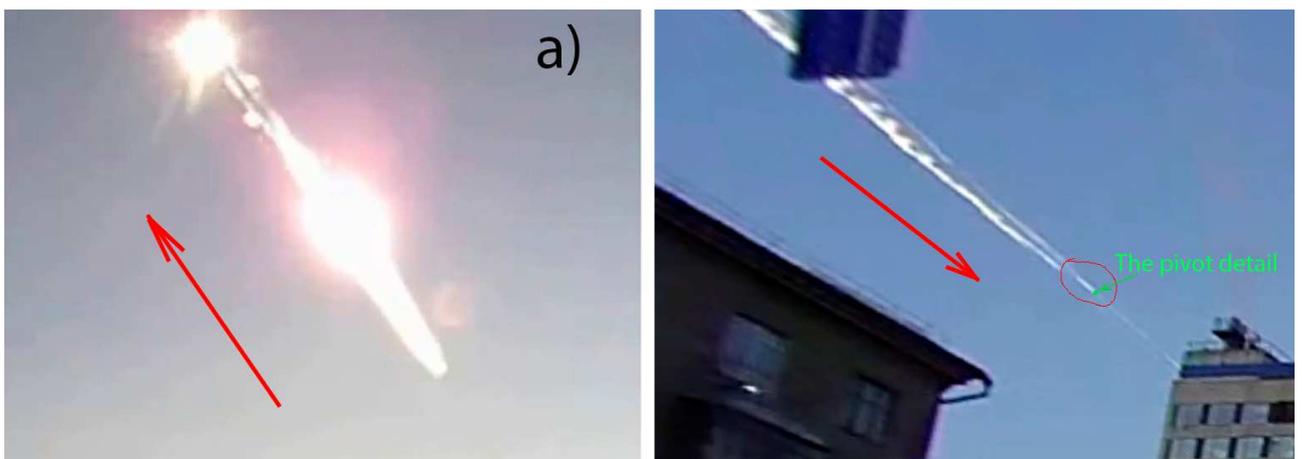

Figure 4. The red arrows show the direction of the movement, a) the dispersion of the fragments on the video record that is made near Chebarkul on M5 road [3]. b) The position of the pivot on the trace [4].

Then the attention will be given to the base fragment, the glow of which is traced for the maximally long time and which presumably fell in Chebarkul with crater forming in the ice. The selected pivot detail of the trace (outlined in the figure 4b) is situated in the district of settlement of Pervomaysk just right after the main flash. In connection with the possibility of alteration of the movement direction of the base fragment, the trajectory in the calculation is divided by two segments: before the pivot – the entry into the atmosphere and after the pivot – the braking and falling. The rotation of these two segments of the trajectory in the iterations of conjunction with the observations is made independently. We should note that the obtained angle 1.6° between the directions of the flight in and out from the pivot point is fairly small and situated primarily in the vertical plane. The azimuth and the altitude of the tangent line to the initial segment are equal to 100.64° and 15.61° with the overall error of the direction of 0.93°. The azimuth and the altitude of the tangent line to the final segment of the trajectory is equal to 280.88° and -17.14° with the overall error of the direction of 1.50°. Quite a raw error in the determination of the direction of the final segment is connected with the unsuccessful position of the fading point of the base fragment on video image and small distance of the fading point from the pivot point.

The input data for the trajectory calculation are:

1) a lot of straight sight lines to the pivot detail and to the fireball at the initial segments of the movement, and at the final segments to the full decline of the observed glow;

2) the azimuths from Snezhinsk to the position of the fireball at two specified time points;

3) the longitude of falling of the base fragment that is equal to the longitude of the ice crater in the lake of Chebarkul.

Google Earth system [5] and specially designed program of the image stacking were used with the compensation of distortion and recalculation of the coordinates of the pixels to the alt-azimuth coordinates in order to obtain the sight lines. A few trips were made to the places of drive recording. The video records were made by CANON 500D master digital camera with the lenses EF-S 18-135 mm and EF-S 15-85 mm during these trips for all video records that are used in the calculations. The records were made from the point where the car was placed at the moment of video recording or in the same manner of movement on the roadway. The specified place for the video recording is the junction in the city of Snezhinsk where the movement of shadows from the vertical arrises can be seen (figure 1). So there were made the pictures of the house against the background of sky of stars (figure 5) that allow to determine the azimuths of the arrises with the precision of minute of arc.

In the simulation the adjusted atmosphere is used that obtained by junction of the standard atmosphere to the upper air sounding data [6] (dependence of the pressure, temperature, density, speed and wind direction on the altitude). The important factors of the movement simulation at the final segment are the ablation and dependence of the drag coefficient on the speed and temperature of the fragment surface. The ablation was taken into account according to the work [7], the work [8] was used for the calculation of the deceleration.

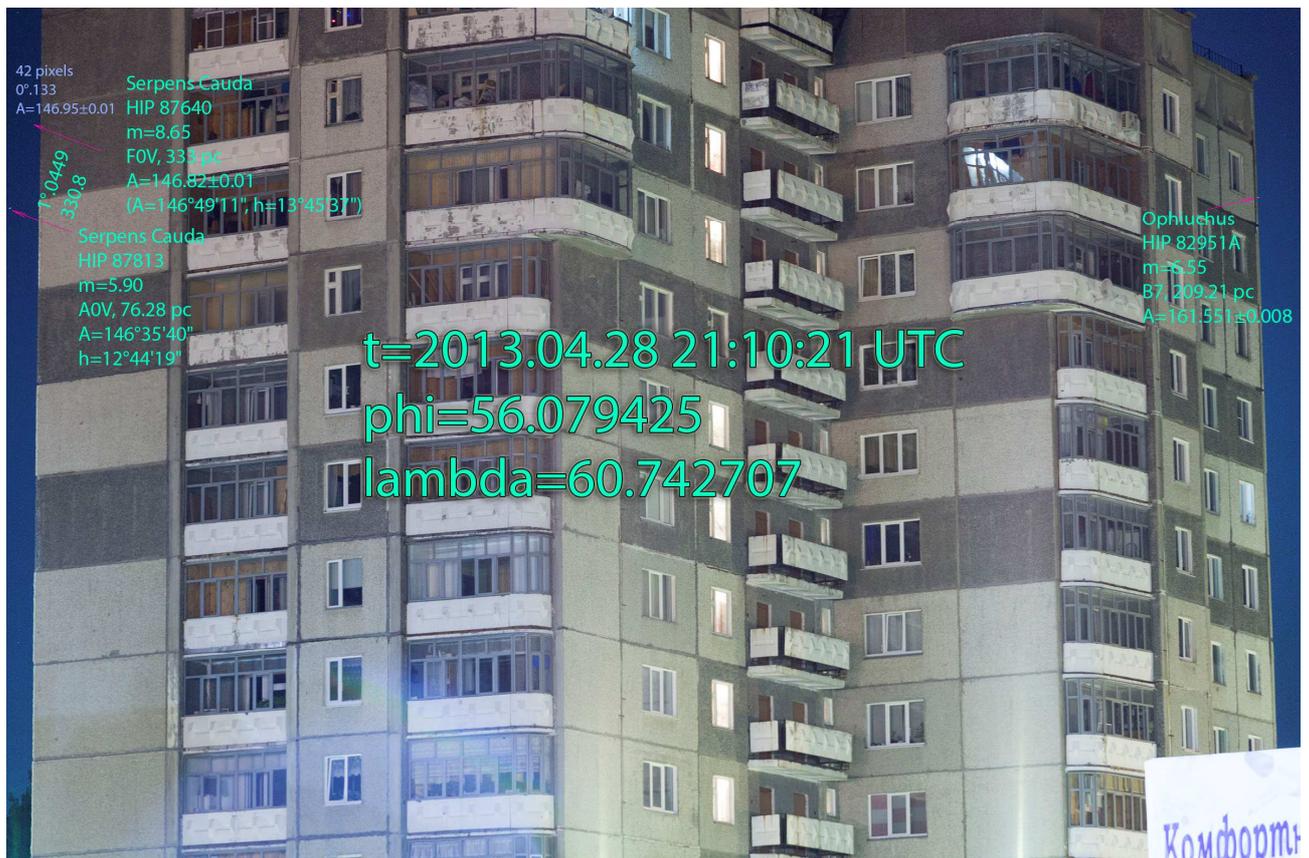

Figure 5. The determination of the azimuth of the arrises by stars HIP 87640 and HIP 82951A. The azimuth of the left arris is equal to the 146.95°±0.01°, the azimuth of the right arris is equal to the 161.551°±0.008°. The distances to the arrises are equal to 143 m and 161 m accordingly.

In the result of the self consistent trajectory simulation, (iterations are converged quickly) the following is determined:

1) The position of the base fragment in the space depending on time.

2) The velocity of the entry into the atmosphere and radiant – point on the celestial sphere where the meteoroid was moving from.

3) Dependences on the time of the velocity modulus, altitude, longitude, latitude, Reynolds numbers, Mach numbers, etc.

4) Dependence on time of the base fragment mass at the final segment of the trajectory.

5) The mass, velocity, angle of falling of the base fragment on the ice of Chebarkul Lake.

The method of calculation of the sight line, self consistent trajectory determination and the results are described further.

The video recorders are equipped with the wide angle lenses that have the larger projection distortion as a rule. Assume the gnomonical projection of the sphere is a picture that is not distorted by the optics (such picture can be obtained in the camera obscura). Let's consider the spherical coordinate system where the optical axis of the lens directed at the pole. All CCD matrices in the video recorders photographic cameras have the square pixels so multiplying the coordinates in pixels by linear size of the pixel we can obtain the coordinates of the image point location. Assume that the plane of the CCD matrix is set perpendicular to the optical axis of the system (but the optical axis does not necessarily go through the center of the frame). Then we can go to the plane polar coordinates on the image. In the result, if there is the gnomonic projection we shall obtain the simple transition from the spherical coordinates $\{\varphi, \theta\}$ to the point position of the frame $\{r, \phi\}$ in the polar coordinate system.

$$\begin{cases} r = f \cdot \tan\theta \\ \phi = \varphi \end{cases}, \qquad (1)$$

where $f$ is the focal distance. During the process of the ray propagation through the lens optical system, the angle $\varphi$ does not change due to the axial symmetry of the optical system and is equal to the polar angle $\phi$. The polar distance $r$ in the picture can change by the optical system (this is a distortion), but as a rule, the function of polar distance recalculation does not change it greatly on all field of shot. So the distortion can be approximated with the sufficient precision by the polynomial as

$$\tilde{r} = \left(1 + ar + br^2\right)r. \qquad (2)$$

Whereas the focal distance of the optical system can be determined by the formula

$$\tilde{f} = \lim_{\theta \to 0} \frac{\tilde{r}}{\theta}. \qquad (3)$$

In the designed image stacking program, the calibration of the master photographic camera and lenses is done first. For this purpose the images of the starry sky that are made by the master photographic camera and the images of the star map with the coordinate grid generated by Stellarium software are align. The signature of the master lens is determined in this calibration that further is used during the measurement of the sight lines. The signature $\{\tilde{f}, a, b, x_0, y_0\}$ includes the focal distance, coefficients of the function of the polar distance recalculation and position of the optical axis on the frame.

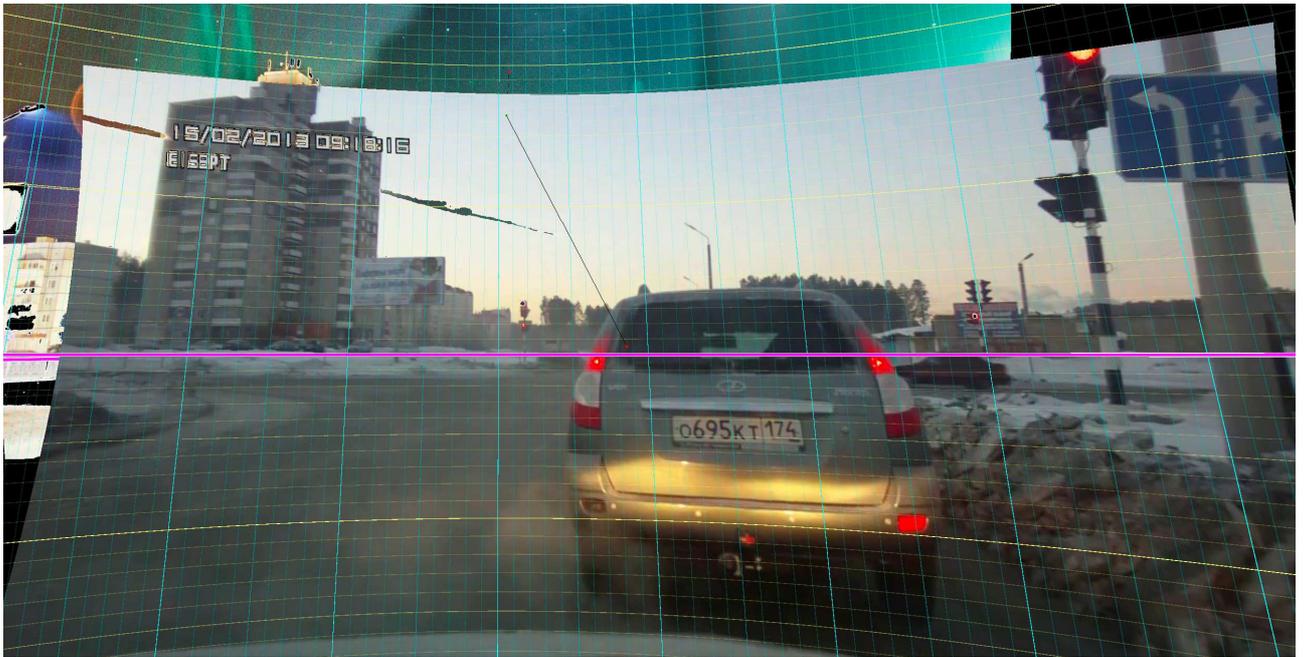

Figure 6. The example of superimposing of the alt-azimuth grid. Firstly, the parameters of the calibration frame with the sky of stars are determined, and then the frame from the video record is set with the calibration frame in the picture outlines, with that the alt-azimuth grid is superimposed.

Further the stacking of the image that was shot by the master photographic camera with the image of the bolide. During the process of stacking, the alt-azimuth coordinates of the optical axes of the images and their positional angles are specified, the signature of the video record is selected at this time that provides the maximum alignment of the images. In the result of such alignment, alt-azimuth grid is set over the frame with the bolide image (figure 6). The coordinates of the optical axis are selected in such a way that the alt-azimuth grid that is superimposed on the calibration frame outlined most accurately the horizon and azimuths of the objects in the picture that were specified in Google Earth system or by coordinates of the celestial bodies (the Sun, the Moon, stars) at the moment of calibration frame.

Such alignment procedure allows to set most accurately the coordinate alt-azimuth grid for the studied frame with the bolide image and determine the directions of the sight lines. The direction to the bolide from Snezhinsk was determined with the error of about 5' (picture 8).The error in this case is determined by the size of the pixels at the original image.

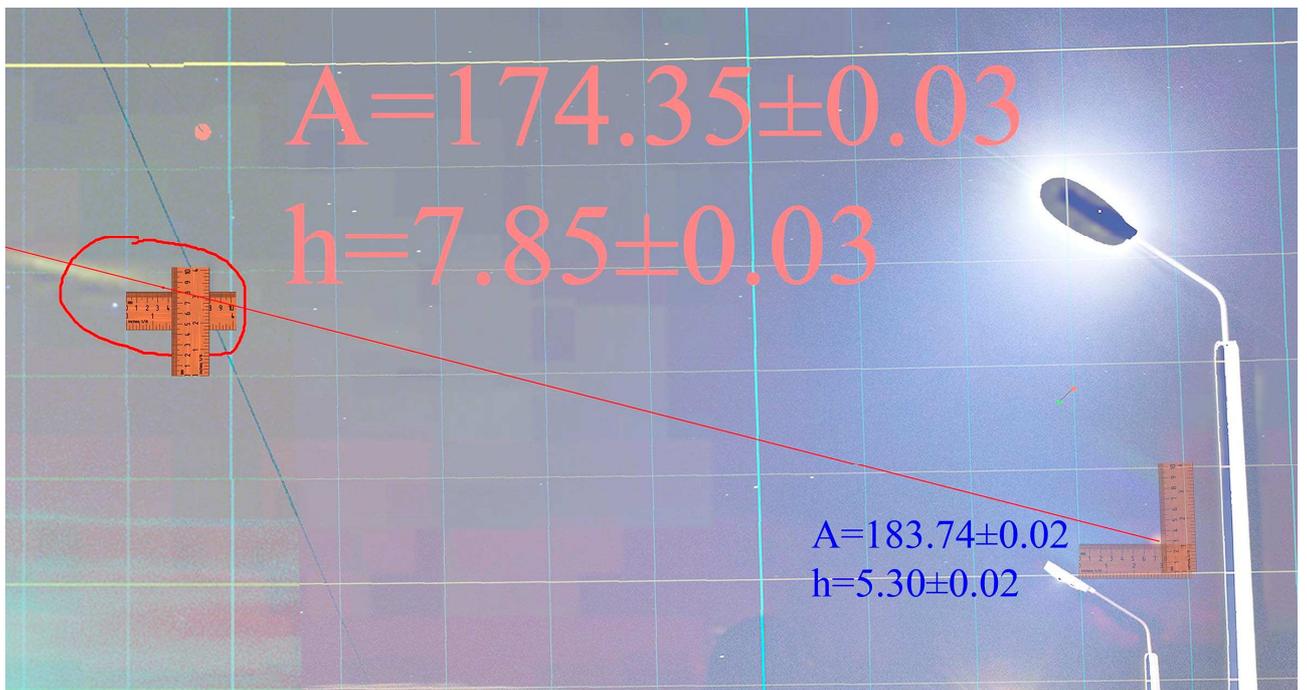

Picture 7. The determination of the pivot point coordinates and luminous fragment before decline. The presence of the light pillars that are situated near allows to determine the positions with the high accuracy. The coordinates of the pivot point are pointed out with the red text; the coordinates of the declining fragment – with the blue text. The errors are specified for the calibration frame. The final error of the determination of the fragment position is about 5' The coordinate grid is set according to the frame of the sky of stars that was made from the position of the video recorder with the error of about 5 cm.

The sight lines of the pivot point with the assessments of the errors by azimuth altitudes are shown in table 1. As we can see, the coordinates of the point are the same as the coordinates on video records with the accuracy of the assessed error.

The latitude, longitude, and altitude of the observation place are set in the system that is used by Google Earth [5]. During the process of recalculation of coordinates in the space and movement simulation of the meteoroid in the atmosphere, the geocentric coordinate system is used with the right basis vector system where OZ axis is directed to the North Pole along the Earth axis and OX axis – to the zero meridian.

A small extended cloud was chosen as the pivot detail, this cloud was observed a moment later after the falling of the base fragment. This characteristic detail that can be found on many video records is difficult to confuse with other details. Whereas it is small enough to determine its coordinates precisely, on the other hand it is big enough to notice it on the video records with the low resolution or made from the long distance.

Table 1. The sight lines of the pivot point (figure 4b) from the different places. If the place of recording is Chelyabinsk then the street only will be specified. The position of shooting site, the measured direction to the pivot point and model direction are presented.

| Site of shooting | Latitude [°] | Longitude [°] | Site height [m] | Measured azimuth [°] | Measured altitude [°] | Model azimuth [°] | Model altitude [°] |
|---|---|---|---|---|---|---|---|
| Yemanzhelinsk Town | 54.756679 | 61.303827 | 235 | 303.5 ± 0.5 | 35.9 ± 0.3 | 303.50 | 35.36 |
| "Park City" Hotel | 55.149930 | 61.363692 | 258 | 223.2 ± 0.5 | 26.5 ± 0.2 | 223.17 | 26.41 |
| Vorovskogo Street | 55.147344 | 61.383465 | 256 | 224.7 ± 1.0 | 26.0 ± 1.0 | 224.79 | 26.02 |
| M5 road | 54.918564 | 60.331521 | 335 | 94.0 ± 0.5 | 26.4 ± 0.3 | 93.94 | 26.11 |
| Rossijskaja Street | 55.160832 | 61.417412 | 237 | 225.6 ± 0.1 | 23.9 ± 0.4 | 225.46 | 24.63 |
| Bejvelja Street | 55.220093 | 61.296243 | 247 | 211.8 ± 0.2 | 24.5 ± 0.2 | 211.68 | 24.44 |
| Kozhzavodskaja Street | 55.191873 | 61.392893 | 216 | 220.7 ± 1.0 | 23.9 ± 1.0 | 220.79 | 23.92 |
| The first Five-Year Street | 55.166315 | 61.444750 | 231 | 226.5 ± 0.5 | 23.6 ± 0.5 | 226.51 | 23.79 |
| Miass Town | 55.105930 | 60.127835 | 333 | 114.1 ± 1.0 | 18.3 ± 0.5 | 114.06 | 18.50 |
| Troitsk Town | 54.077058 | 61.531300 | 183 | 337.9 ± 0.2 | 11.0 ± 0.2 | 337.56 | 10.87 |
| Snezhinsk Town | 56.079425 | 60.742707 | 265 | 174.3 ± 0.2 | 7.8 ± 0.2 | 174.36 | 7.80 |
| Kamensk-Uralsky Town | 56.415199 | 61.918560 | 170 | 200.4 ± 1.0 | 5.4 ± 0.5 | 200.26 | 5.42 |

In order to find the position of the pivot detail we shall find the point in space that would be close to all sight lines that are run in the direction of the pivot detail of a trace. The simple numerical algorithm was used; the sample point is moved for a small distance in the direction of the nearest point on the sight line. The displacing is performed in turns in the direction of all sight lines. The moving to the sight lines is multiply repeated till the moment when the sample point stops drifting in space. The determined position is shown in the figure 8 together with the sight lines. The mean-square distance of the point from the sight lines is 320 m. The height of the supporting point above the ground surface is 19.6 km, the latitude is 54°.8926 and the longitude is 60°.9461.

Let's consider the meteoroid trajectory in the atmosphere in general. The impacts of atmosphere deceleration influence at the initial section of meteoroid movement are negligible. As a result, till the main flash, i.e., till the moment of fireball emerging from behind the building on the video record [2] (figure 1), the meteoroid movement was accelerated in the gravitational field of the Earth without noticeable decelerating in the atmosphere. Due to the short period of this movement phase and to the fact that the force of gravity is almost orthogonal to the movement direction, the velocity modulus at the initial phase of the movement can be considered as the constant one with the fine precision.

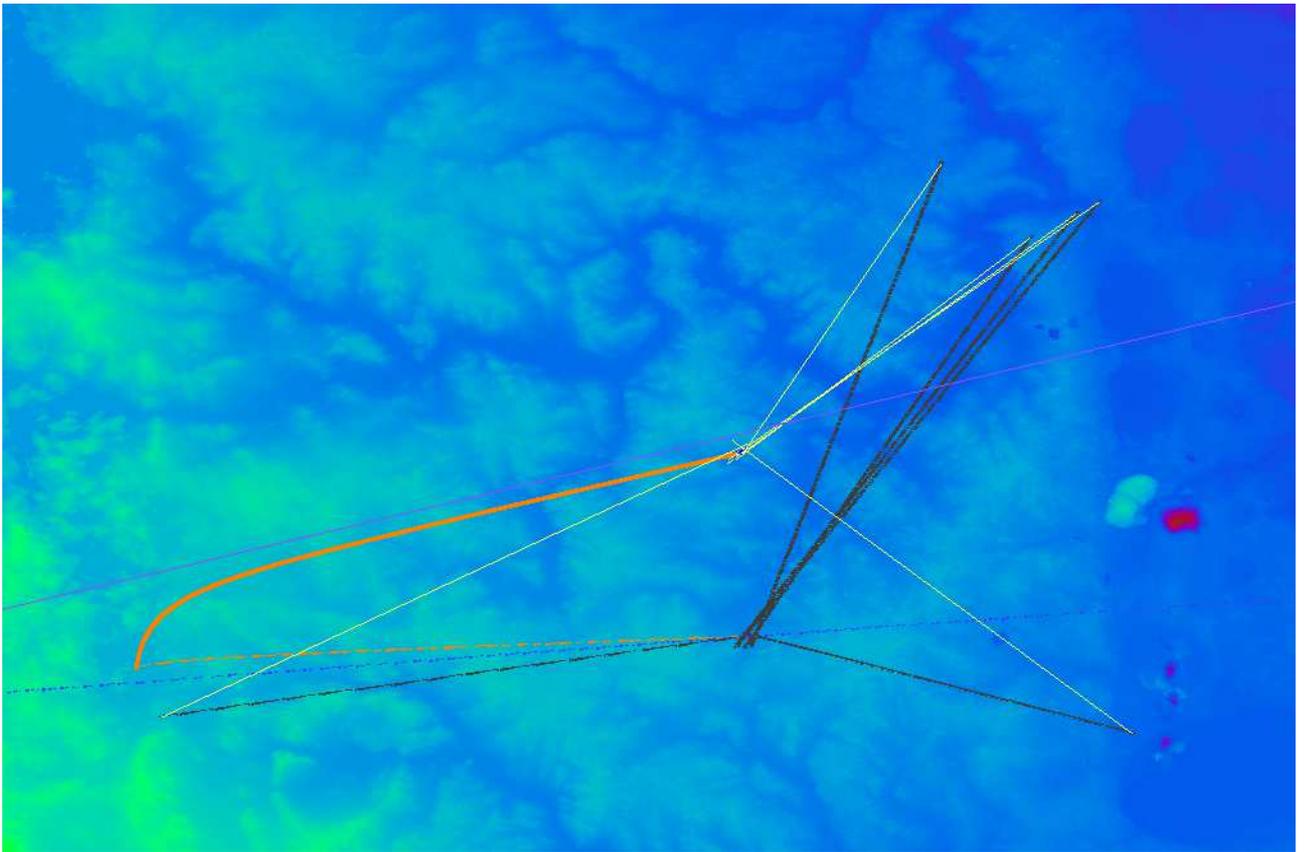

Figure 8. The three-dimensional trajectory model. The color of the point on the ground surface is determined by its height above geoid according to the satellite radiolocation data [9]. Korkinskiy opencast coal mine is seen as the magenta spot on the right. The orange curve shows the trajectory of Chebarkul meteorite, the contours of Chebarkul Lake are seen, a place where it fell. The sight lines to the pivot point are shown in yellow color, their projections on the ground surface are black.

Quite different picture is seen at the final segment of movement of the base fragment. It was many times as small of the original body and moved in the dense atmosphere. As a result, the strong ablation and the rapidly increasing of the braking acceleration affected the fragment till it reached the speed of sound at a height of about 10 km. After that, the sound wave got off the fragment, the coefficient of drag fell abruptly and the movement quickly evolved into the free vertical fall. From the video record that was presented by Nikolay Lavrentievich Melnikov (figure 9) we can obtain not only the moment of falling in lake (62.4 seconds after the maximum of glowing) but also the moment of the coming sound wave that preceded it (46.8 seconds). These moments are well correlated with the results of simulation (61.1 s for the moment of falling and 47.6 s for the moment of the sound coming).

The calculation of ablation is performed according to the work [7]. The flying range depends on initial mass of the fragment and on the drag coefficient change along trajectory for the spherical body. Notably, the flying range is weakly dependent on the mass of body and strongly dependent on drag coefficient under hypersonic speeds. So the drag coefficient that was obtained with high accuracy shall be used in the movement simulation of the base fragment. The dependences of the drag coefficient on the Reynolds, Mach and surface temperature were obtained from the work [8]. The main factor of simulation of the final trajectory segment is the state of the atmosphere as well. The analysis of the pressure profiles was performed including temperature and wind that were obtained at the stations of upper air sounding. The analysis showed up the similarity of wind profiles in the region behind the Ural mountain range. Thus, the data from Verkhneye Dubrovo that were obtained in 3 hours before the event can be considered as the data that describe the state of the atmosphere in the most precise way. The data were merged with the modified standard atmosphere at the height of more than 30000 m. The pressure and temperature profiles are shown in the figure 10. The dependence of the wind speed on altitude is represented in the figure 11.

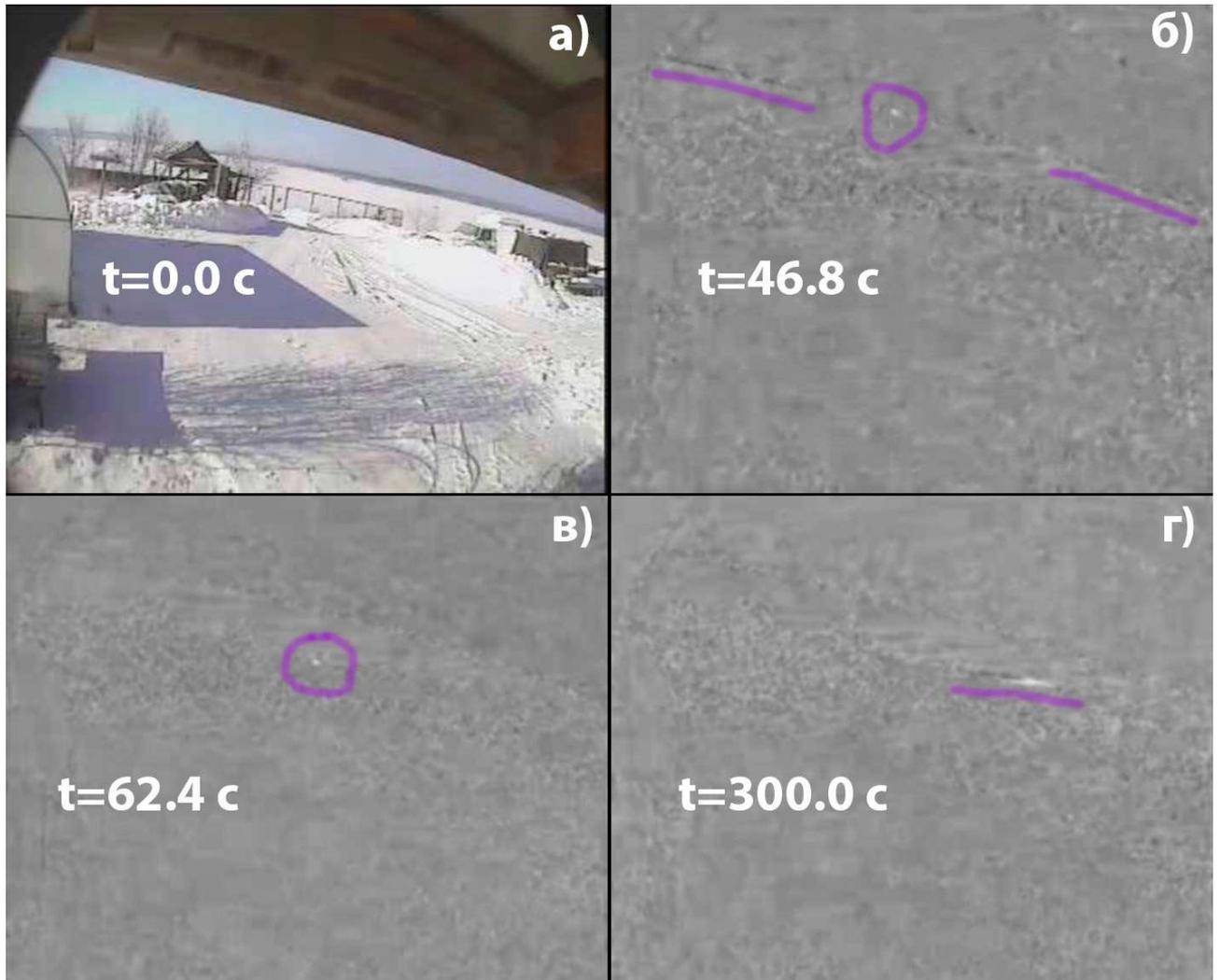

Figure 9. The falling of Chebarkul meteorite according to the video record of Nikolay Melnikov. In figure a) the frame from video record is shown, in figures b)-d) the processed differences between the current frame and average from 100 of preceding frames. a) The moment of the highest brightness of snow that is lightened by Chelyabinsk bolide. b) The coming of the sound wave leads to the movement of the edge of the oil cloth that is hanged down the roof. At this moment the nubbin of snow falls off the oil cloth. c) The falling of the meteorite with the creation of the snow-and-ice plume. d) The drift of tail of the snow dust under the influence of south-east wind.

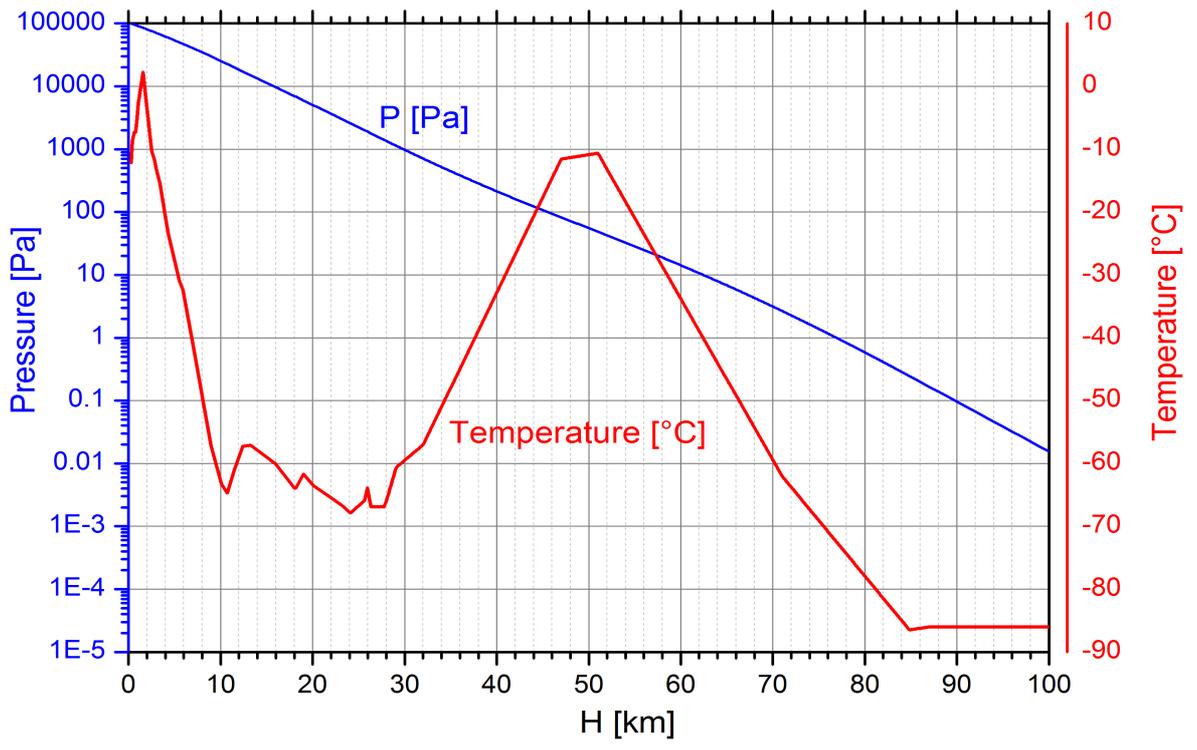

Figure 10. The dependence of pressure and temperature on altitude. The data from Verkhneye Dubrovo are used below 30 km (see figure 4) The standard atmosphere that is modified by temperature for merging with the observed data is used for the upper atmosphere.

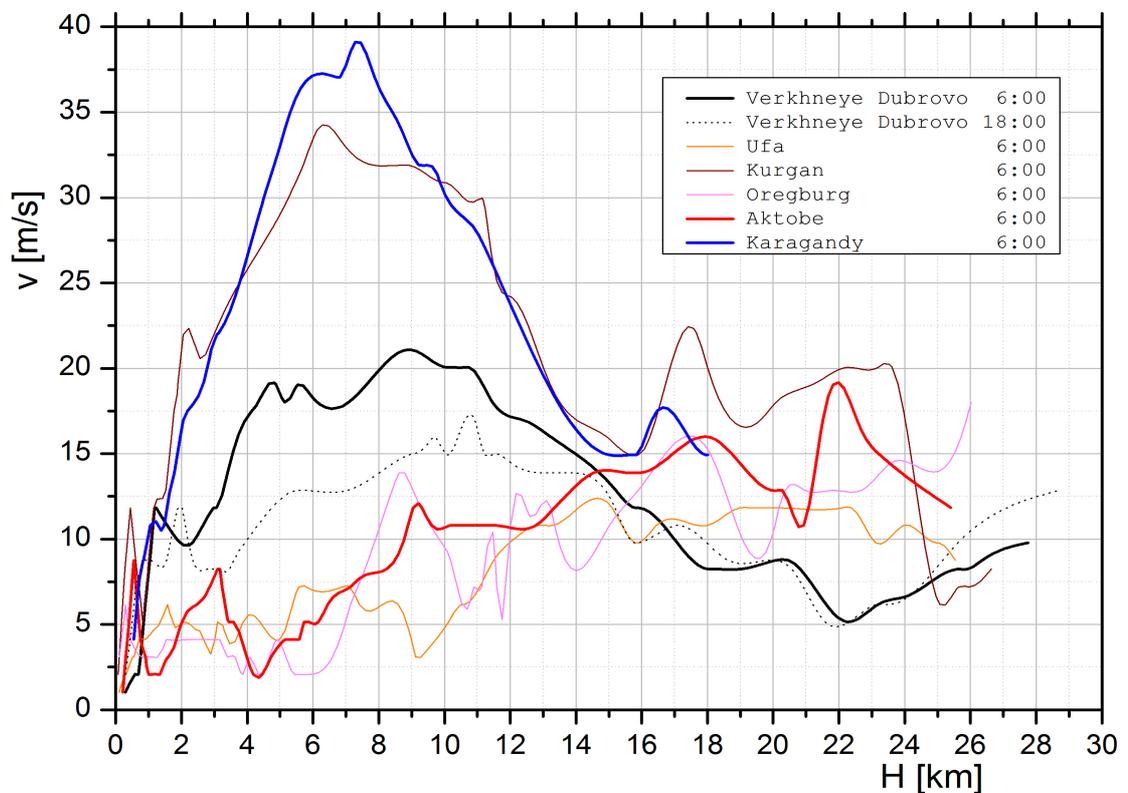

Figure 11. The dependence of the wind velocity on altitude in the different points of observation. In the east (Kurgan, Karaganda) the wind is significantly stronger than behind the Ural mountain range. Verkhneye Dubrovo takes the intermediate position, whereas the wind grows weak to the evening on February 15. The profile from Verkhneye Dubrovo on 6 o'clock was used for the calculation. The wind direction is almost constant (330°, i.e., north-west wind) except the lower layer that is about 1 km.

The movement trajectory was simulated by the method of Verlet with the time step of 0.01 s. The calculation is performed in the inertial Cartesian coordinate system that is moving with the speed of the surface in the region of Chelyabinsk. The mass evolved due to the ablation, the fragmentation was not included because it is not important for determination of the trajectory (the movement of the base fragment in separate or along with the rest mass till the moment of the main flash are almost similar due to rare atmosphere). The variable gravitational field and the atmosphere deceleration force are taken into account.

The self-consistent trajectory calculation includes the following stages
1) The optimization of the flying range of the base fragment using the change of mass and velocity modulus in the pivot point of the trajectory if the speed of entry into the atmosphere is similar to the speed specified in 5).
2) The determination of points on the trajectory that are nearest to the sight lines of the bolide at the initial and final segment.
3) The turn of the trajectory with the stationary pivot point that lead the trajectory to the sight lines.
4) The determination of the points on the trajectory that are corresponding to the azimuth of the arrises in Snezhinsk.
5) The determination of the velocity of entry into the atmosphere using the adjustment of time of movement between the discovered points of the trajectory to the observable time of flying between them in 2.84±0.01 s.

In the beginning of iteration, the mass of the base fragment in the pivot point is specified equal to 10000 kg, the direction of movement is strictly to the east horizontally with the velocity of 10 km/s. The self-consistent loop converges rapidly and by several iterations the trajectory stops to change. The calculation of the trajectory is performed in the internal optimization cycle. The trajectory is calculated from the pivot point to the place of falling forward in time and then the back-calculating of the initial segment back in time is performed (integration of the movement up to -30 s regarding the pivot point). The input data for such calculation of the trajectory is a mass of the base fragment in the pivot point, its velocity modulus and two direction vectors for the initial and final segment of the trajectory.

The simulation found that:
1) The body entered the atmosphere at the speed of 19.01±0.1 km/s.
2) The acceleration of braking reached 1.0g at the height of 43 km. Thus, even if the body was not solid but represented a mass of rocks of different size down to the dust fraction (rubble pile), the fragmentation of this mass would happen at the height of 40 km that does not contradict with the visible luminous intensity at the different heights.
3) In case of absence of the atmosphere the body would hit the surface at an angle 15.2° of latitude 55.0130° N and longitude 59.8599° E.
4) The simulation of falling of Chebarkul fragment gives mass of 911 kg at the distance of 0.5 km to the south of the ice crater. Notably, this is the sufficient precision in calculation taking into account some accidental indefiniteness in the fragment separation and air gusts. The falling occurred at an angle of 78.6° to the horizon (almost vertically) at the speed of 213 m/s. During such entry into the surface of the frozen lake, the deviation of the meteorite position at the bottom from the centre of the crater shall be about 1.5 m.

The projection of the trajectory on the ground surface is in the north of the main part of the discovered meteorites (figure 12), that confirms the drifts of the meteorites by the wind during the falling in south-west. The lesser mass of the meteorite is, the earlier it brakes, the higher altitude it falls off and lower speed it has. The wind drift grows as a result with the increasing of longitude.

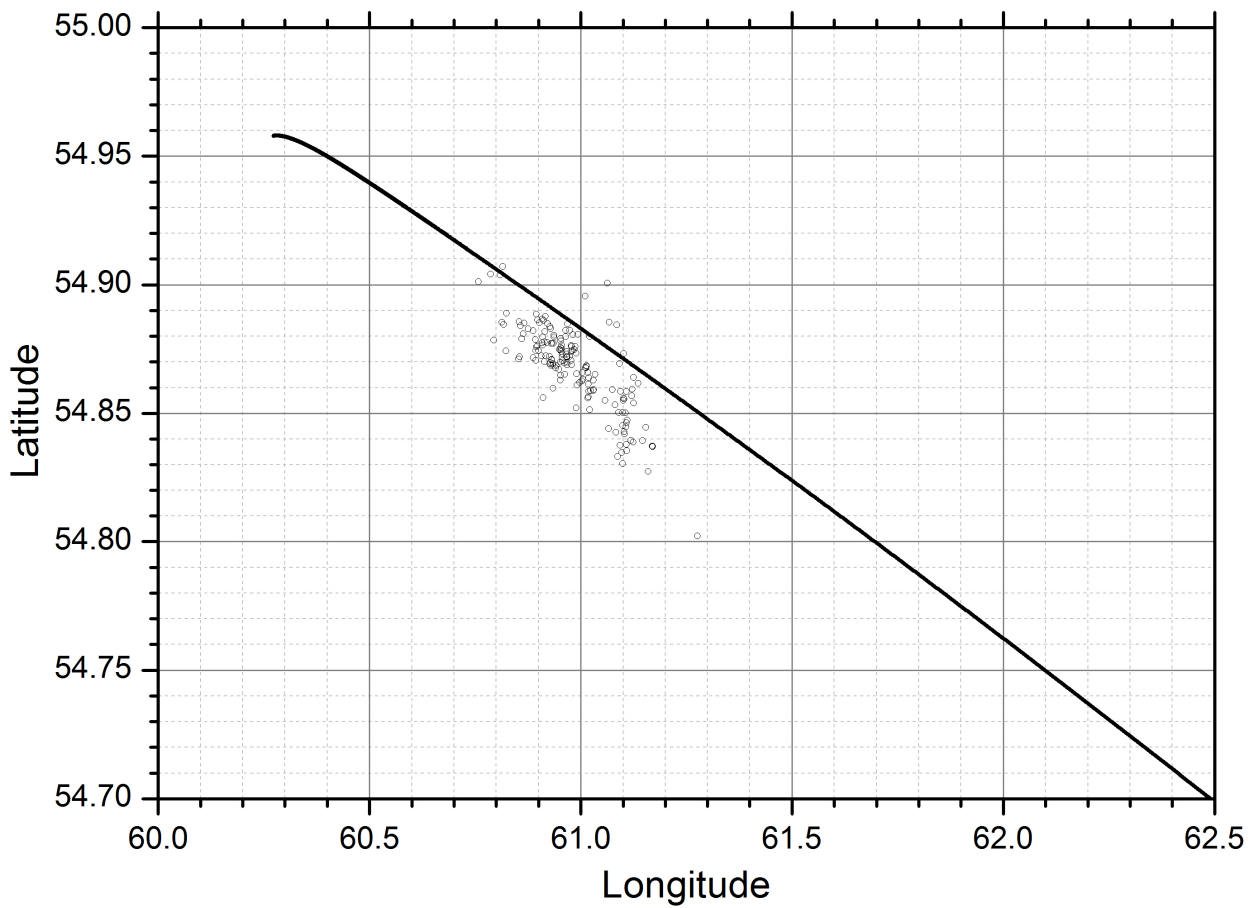

Figure 12. The projection on the ground surface of the calculated movement trajectory of Chebarkul meteorite. The circle shows the known discovery places of meteorites. The falling of the meteorites to the south of the trajectory occurred due to wind drift during the steady falling.

The trajectories of the fragments are converged to the extreme trajectory in case of mass increasing and mass fragment of about 10 ton the period of falling is weakly dependent on the degree of its atmospherical braking. The acceleration of the fragment during its braking is proportional to the force and inversely proportional to the mass but the force is proportional to the cross-section that is proportional to the mass raised to the power of 2/3 in the case of spherical body. As a result, the twofold error in the drag coefficient will lead to the eight-fold error in the determination of mass of the body that fell into the specified point. So the calculated mass of Chebarkul fragment in 911 kg (that is close to the total mass of the fragments that were gathered from the bottom of Chebarkul Lake taking into account the fragments that were not found) tells us about precision accuracy in determination of trajectory of entry into the atmosphere and high quality of the used structural models of the atmosphere, ablation and drag coefficient.

It is shown in the figure 13 that during the movement (in the side of longitude decreasing) that the change occurred in the atmosphere density, the altitude of base fragment above the surface of Earth, Mach number and the angle of movement regarding horizon. The acceleration that affected Chebarkul fragment reached 300 g at the height of 20 km. At this time its mass was 2600 kg and the power of transition of kinetic energy in some other forms was about 300 GW (350 units of nuclear facility).

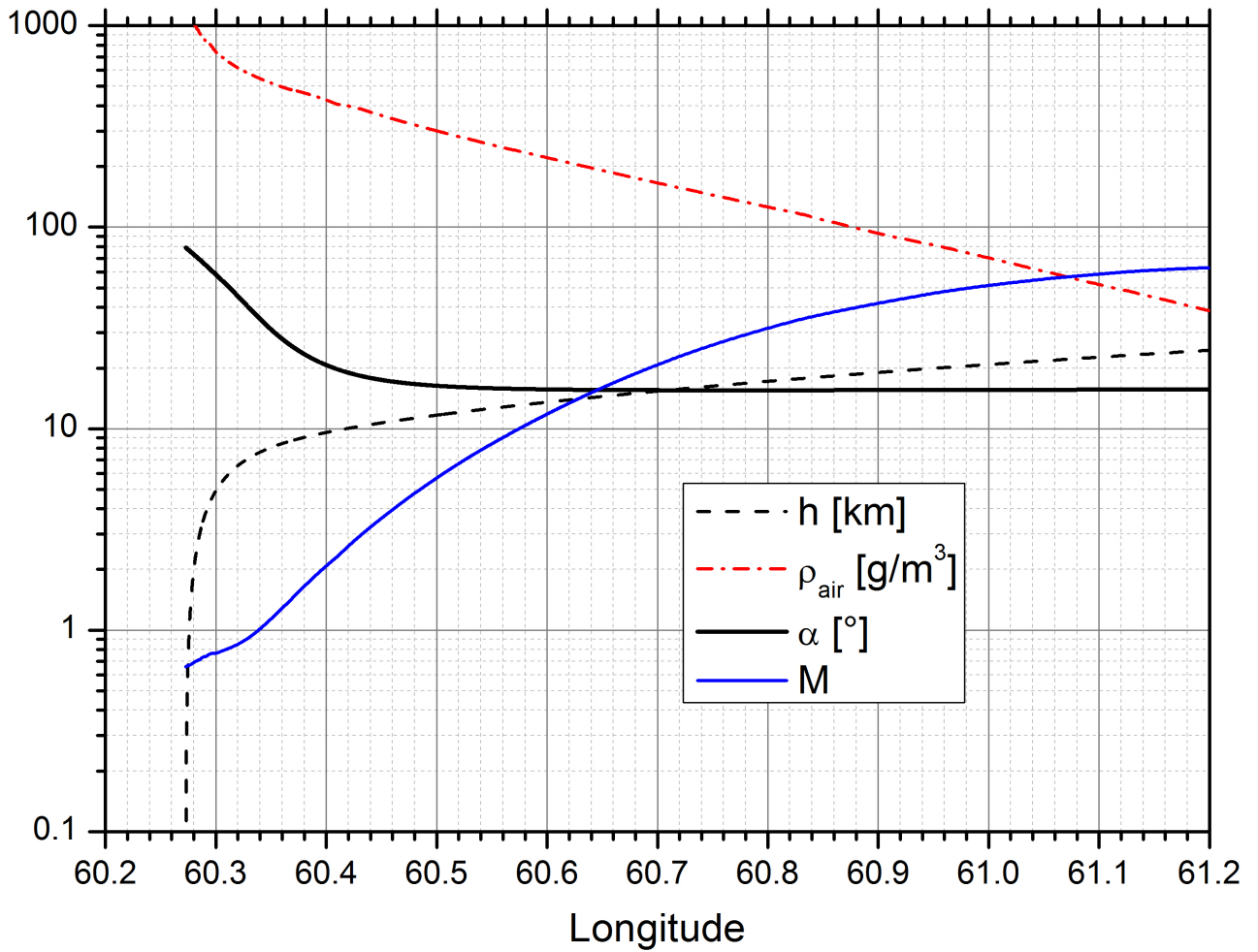

Figure 13. The dependences of altitude, vertical angle, Mach number and air density on the longitude along the trajectory of Chebarkul meteorite. The step-like data in the diagram of the vertical angle is connected with the transition from the initial segment of the trajectory to the final segment that were simulated as independent.